\documentstyle[aps,prd, preprint]{revtex}

\begin{document}
\title{Non-relativistic quantum systems on topological defects space-times}
\author{Geusa de A. Marques and Valdir B. Bezerra.}
\maketitle

\begin{center}
{\it {Departamento de F\'{\i}sica-CCEN, Universidade Federal da Para\'{\i}ba,%
}}

{\it {Caixa Postal 5008, 58051-970, Jo\~{a}o Pessoa, Pb, Brazil.}}
\end{center}

\abstract

We study the behavior of non-relativistic quantum particles
interacting with different potentials in
the space-times generated by a cosmic string and a global monopole. We find
the energy spectra in the presence of these topological defects
and show how they differ from their free-space-time values.


\vskip 3.0 cm \centerline{\bf{I. Introduction} }

The study of quantum systems under the influence of a gravitational field
has been an exciting research field. Along this line of research the
hydrogen atom, for example, has been studied in particular curved
space-times\cite{1,2}. An atom placed in a gravitational
field will be influenced by its interaction with the local curvature as well
as with the topology of the space-time. As a result of this interaction, an
observer at rest with respect to the atom would see a change in its spectrum.
This shift in the energy of each
atomic level would depend on the features of the space-time. The problem of
finding these shifts\cite{3} in the energy levels under the influence of a
gravitational field is of considerable theoretical interest as well as
possible observational. These shifts in the energy spectrum due to the
gravitational field are different from the ones produced by the
electromagnetic field present, for example, near white dwarfs and neutron
stars. In fact, it was already shown that in the Schwarzschild geometry,
the level spacing of the gravitational effect is different from that of the
well-known first order Stark and Zeeman effects, and therefore, in
principle, it would be possible to separate the electromagnetic and
gravitational perturbations of the spectrum\cite{3}. Other investigation
concerning this interesting subject is to use loosely bound Rydberg atoms in
curved space-time to study the gravitational shift in the energy
spectrum\cite{4}.

The first experiment which showed the gravitational effect on a wave
function was performed by Colella, Overhauser and Werner\cite{5a} in which
they measured the quantum mechanical phase difference of two neutron beams
induced by a gravitational field. Another gravitational effect that appears
in quantum interference is the neutrino oscillations\cite{6a} which has been
discussed recently.

The general theory of relativity, as a metric theory, predict that
gravitation is manifested as curvature of space-time. This curvature is
characterized by the Riemann tensor $R_{\beta \gamma \delta }^{\alpha }$.
It is of interest to know how the curvature of space-time at the position of
the atom affects its spectrum. On the other hand, we know that there are
connections between topological properties of the space and local physical
laws in such a way that the local intrinsic geometry of the space is not
sufficient to describe completely the physics of a given system. Therefore,
it is also important to investigate the role played by a nontrivial
topology, for example, on a quantum system. As examples of these
investigations we can mention the calculation of the topological scattering
amplitude in the context of quantum mechanics on a cone\cite{5} and the
interaction of a quantum system with conical singularities\cite{6,6b}.

Therefore, the problem of finding how the energy spectrum of an atom placed
in a gravitational field is perturbed by this background has to take into
account the geometrical and topological features of the space-time under
consideration and in this way we should emphasize that when a quantum system
is embedded in a curved space-time it is influenced by its structure and
topology. In other words, the dynamic of atomic systems is determined by the
curvature at the position of the atom and also by the topology of the
background space-time.

According to standard quantum mechanics, the motion of a charged particle
can be influenced by electromagnetic fields in regions from which the
particle is rigorously excluded\cite{10a}. In this region the
electromagnetic field vanishes. This phenomenon has come to be called
Aharonov-Bohm effect after a seminal paper by Aharonov and Bohm\cite{10a}.
It was shown that in the quantum scattering problem the background leads to
a non-trivial scattering, which was confirmed experimentally\cite{11a}.

The analogue of the electromagnetic Aharonov-Bohm effect set up is the
background space-time of a cosmic string\cite{7} in which the geometry is
flat everywhere apart from a symmetry axis.

Cosmic strings\cite{7} and monopoles\cite{8} are exotic
topological defects\cite{9} which may have been formed at phase
transitions in the very early
history of the Universe. Up to the moment no direct observational evidence
of their existence has been found, but the richness of the new ideas they
brought along to general relativity seems to justify the interest in the
study of these structures.

The gravitational field of a cosmic string is quite remarkable; a particle
placed at rest around a straight, infinite, static string will not be
attracted to it; there is no local gravity. The space-time around a cosmic
string is locally flat but not globally. The external gravitational field
due to a cosmic string may be approximately described by a commonly called
conical geometry. Due to this conical geometry a cosmic string can induce
several effects like, for example, gravitational leasing \cite{10}, pair
production\cite{11}, electrostatic self-force\cite{12} on
an electric charge at rest, bremsstrahlung process\cite{13} and the so-called
gravitational Aharonov-Bohm effect\cite{14}.

The space-time of a point global monopole has also some unusual properties.
It possesses a deficit solid angle $\Delta =32\pi ^{2}G\eta ^{2}$, $\eta $
being the energy scale of symmetry breaking. Test particles in this
space-time experiences a topological scattering by an angle $\pi \frac{\Delta
}{2}$ irrespective of their velocity and their impact
parameter. The effects
produced by the point global monopole are due to the deficit solid angle
which determines the curvature and the topological features of this
space-time.

The aim of this paper is to study the energy shift associated with a
non relativistic quantum particle interacting with different
potentials in the space-times generated by a cosmic string and a global
monopole.

In the cosmic string case, we consider the harmonic oscillator
and the Coulomb potentials and determine how the nontrivial topology of
this background space-time perturbs the energy spectrum. The influence of the
conical geometry on the energy eigenvalues mani
fests as a kind of
gravitational Aharonov-Bohm effect for bound states\cite{6,19a}, whose
analogue in the electromagnetic case shows that\cite{20a} the bound
state energy depend on the external magnetic flux in a region from which
the electron is excluded.

In the case of a point global monopole we are concerned with a  similar
proposal. We will investigate the interactions of a non-relativistic quantum
particle with the Kratzer\cite{15} and the Morse\cite{16} molecular potentials
 in this background  space-time. In this case we also determine the shifts
in the energy levels.

These modifications in the energy spectra as compared to the simplest
situation of empty flat Minkowski space-time, could be used, in principle, as
a probe of the presence of these defects in the cosmos. The possibility of
using this effect for such a purpose is interesting because at present our
observational knowledge of the existence of such defects is quite limited
and this effect offers one more possibility to detect these topological
defects. In fact to produce observable modifications in the spectra from
the astrophysical point of view, one needs an huge number of
 particles in the states we are studying, otherwise the
 contribution to real spectral effects is thus not
 likely to be sufficiently strong to be observed.

In order to do these studies let us consider that a non-relativistic
particle living in a curved space-time is described by the Schr\"{o}dinger
equation which should take the form
\begin{equation}
i\hbar \frac{\partial \psi }{\partial t}=-\frac{\hbar ^{2}}{2\mu }\nabla
_{LB}^{2}\psi +V \psi ,  \label{a1}
\end{equation}
where $\nabla _{LB}^{2}$ is the Laplace-Beltrami operator the covariant
version of the Laplacian given by
 $\nabla _{LB}^{2}=g^{-\frac{1}{2}}\partial
_{i}\left( g^{ij}g^{\frac{1}{2}}\partial _{j}\right) $,
 with $i,j=1,2,3;$ $%
g=\det \left( g_{ij}\right) ;$ $\mu $
 is the mass of the particle and $V$ is
an external potential. Throughout this paper
 we will use units in which $c=1$%
.

\vskip 1.0 cm

\centerline{\bf{II. Coulomb potential in the space-time of a cosmic string}}

In what follows we analyze the energy level of a particle in the
 presence of a Coulomb potential in the space-time of a cosmic
 string. To do this let us
consider the exterior metric of an infinitely long straight and static
string in spherical coordinates. It reads as
\begin{equation}
ds^{2}=-dt^{2}+dr^{2}+r^{2}d\theta ^{2}+\alpha ^{2}r^{2}\sin ^{2}\theta
d\varphi ^{2},  \label{21}
\end{equation}
with $0<r<\infty $, $0<\theta <\pi $, $0\leq \varphi \leq 2\pi $. The
parameter $\alpha
=1-4G\bar{\mu}$ runs in the interval $\left[ 0,1\right] $, $\bar{\mu}$ being
the linear mass density of the cosmic string. In the
special case $\alpha =1$ we obtain the Minkowski space in spherical
coordinates.

The time-independent Schr\"{o}dinger equation in this case is
\begin{eqnarray}
&&\left. -\frac{\hbar ^{2}}{2\mu r^{2}}\left[ 2r\frac{\partial }{\partial
r}+r^{2}\frac{\partial ^{2}}{\partial r^{2}}+\cot \theta \frac{\partial
}{\partial \theta }+\frac{\partial ^{2}}{\partial \theta ^{2}}+\frac{1}{%
\alpha ^{2}\sin ^{2}\theta }\frac{\partial ^{2}}{\partial \varphi ^{2}}%
\right] \psi \left( r,\theta ,\varphi \right) \right.  \nonumber \\
&&\left. V(r)\psi \left( r,\theta ,\varphi \right) =E\psi \left( r,\theta
,\varphi \right) \right. .  \label{22}
\end{eqnarray}
Let us assume that the eigenfunctions have the form

\begin{equation}
\psi (r,\theta ,\varphi )=\frac{1}{\sqrt{2\pi }}e^{im\varphi }R(r)\Theta
(\theta )  \label{23}
\end{equation}
then, we get the following equations for the radial and angular parts

\begin{equation}
-\frac{\hbar ^{2}}{2\mu }\frac{d ^{2}u(r)}{d r^{2}}%
+V(r)_{eff}u(r)=Eu(r),  \label{27}
\end{equation}
where we changed the function $R(r)$ by $u(r)$ introducing $ u(r)=rR(r)$
and $V_{eff}(r)$ is the effective potential experienced by the particle and
given by
\begin{equation}
V_{eff}(r)=V(r)-\frac{\lambda }{r^{2}}.  \label{28}
\end{equation}
and

\begin{equation}
\left( 1-\xi ^{2}\right) \frac{d^{2}F(\xi )}
{d\xi ^{2}}-2\xi \frac{dF(\xi )}{%
d\xi }-\left( \frac{2\mu \lambda }{\hbar ^{2}}+\frac{m^{2}}{\alpha
^{2}\left( 1-\xi ^{2}\right) }\right) F(\xi )=0,  \label{32}
\end{equation}
where we introduced the change of variables $\xi =
 \cos \theta$ and $ F(\xi )\equiv \Theta (\theta)$.
Equation (\ref{32}) is the generalized associated Legendre equation. It
becomes an equation with eigenvalue $\frac{2\mu \lambda }{\hbar ^{2}}$, if
we demand that the solution be finite at the singular
 points $\xi =\pm 1.$ Its examination follows a conventional pattern.

A convenient method to obtain the solutions of Eq. (\ref{32}) is by the
investigation of its behavior at the singular points $\xi =\pm 1$.
Doing this, the accepted solution may have the form
\begin{equation}
F(\xi )=\left( 1-\xi ^{2}\right) ^{\frac{m}{2\alpha }}G(\xi ),  \label{37}
\end{equation}
where $G$ is analytic in all space except
 when $\xi\rightarrow \pm \infty ,$ and
is different from zero for $\xi =\pm 1.$

From Eqs. ($\ref{32}$) and (\ref{37}), we get
\begin{equation}
\left( 1-\xi ^{2}\right) \frac{d^{2}G(\xi )}{d\xi ^{2}}-2(m_{(\alpha )}+1)\xi
\frac{dG(\xi )}{d\xi }-\left( m_{(\alpha )}^{2}+m_{(\alpha )}
+\bar{\lambda}\right) G(\xi
)=0,  \label{38}
\end{equation}
where
\[
m_{(\alpha )}=\frac{m}{\alpha };\text{ }\bar{\lambda}=
\frac{2\mu \lambda }{\hbar
^{2}}.
\]
Expanding the regular solution of Eq. (\ref{38}) in a power series we get
the recursion relation
\begin{equation}
a_{n+2}=\frac{n\left( n-1\right) +2\left( \bar{m}+1\right)
 n+\bar{\lambda}+%
\bar{m}\left( \bar{m}+1\right) }
{\left( n+1\right) \left( n+2\right) }a_{n},
\label{39}
 \end{equation}
where $n$ is an integer $\geq 0,$ and $G(\xi )$ diverges at $\left| \xi
\right| =1.$ In order to have acceptable eigenfunctions, the series must be
terminate at some finite value of $n$. According to Eq. (\ref{39}), this
will happen if $\bar{\lambda}$ has the value
\begin{equation}
\left. \Rightarrow \bar{\lambda}=-l_{(\alpha )}\left( l_{(\alpha )}+1\right)
\right. ,  \label{40}
\end{equation}
where
\[l_{(\alpha )}=m_{(\alpha )}+n. \]
Therefore, Eq. (\ref{32}) turns into
\begin{equation}
\left( 1-\xi ^{2}\right) \frac{d^{2}F(\xi )}{d\xi ^{2}}-2\xi
 \frac{dF(\xi )}{%
d\xi }+l_{(\alpha )}(l_{(\alpha )}+1)F(\xi )-\frac{m^{2}}{\alpha ^{2}(1-\xi
^{2})}F(\xi )=0,  \label{41}
\end{equation}
which corresponds to a generalized Legendre
 equation in the sense that now $ l_{(\alpha)}$ and
 $ m_{(\alpha)}$ are not necessarily integers. Its solutions are thus given by
\begin{equation}
F_{l_{(\alpha )}}^{m_{(\alpha )}}(\xi )(\xi )=P_{l_{(\alpha )}}^{m_{(\alpha
)}}(\xi )=\frac{1}{2^{l_{(\alpha )}}l_{(\alpha )}!}\left( 1-\xi ^{2}\right)
^{\frac{m_{(\alpha )}}{2}}\frac{d^{m_{(\alpha )}+l_{(\alpha )}}}{d\xi
^{m_{(\alpha )}+l_{(\alpha )}}}\left[ (\xi ^{2}-1)^{l_{(\alpha )}}\right] .
\label{42}
\end{equation}

Now, substituting Eq. (\ref{40}) into (\ref{27}) and considering $V(r)=-%
\frac{k}{r}$, we find
\begin{equation}
\frac{d^{2}u(r)}{dr^{2}}+2k\frac{\mu }{r\hbar ^{2}}u(r)-\bar{\beta}^{2}u(r)-%
\frac{1}{r^{2}}\left[ l_{(\alpha )}(l_{(\alpha )}+1)\right] u(r)=0,
\label{43}
\end{equation}
where
\begin{equation}
\bar{\beta}^{2}=-\frac{2\mu E_{n_{r}}}{\hbar ^{2}};\text{ }E_{n_{r}}<0.
\label{g}
\end{equation}
Equation (\ref{43}) is a confluent hypergeometric
equation whose solution is given by
\begin{equation}
u(r)=_{1}F_{1}\left( l_{(\alpha )}+1-\frac{k^{2}\mu }{\bar{\beta}\hbar ^{2}}%
,2+2l_{(\alpha )};2\bar{\beta}r\right) .  \label{44}
\end{equation}
This function is divergent, unless
\begin{equation}
1+l_{(\alpha )}-\frac{\mu k^{2}}{\bar{\beta}\hbar ^{2}}=-n_{r};\text{ }%
n_{r}=0,1,2...\text{ .}  \label{45}
\end{equation}
Then, from the previous condition we find the energy eigenvalues
\begin{equation}
E_{n_{r}}=-\frac{\mu k^{2}}{2\hbar ^{2}}\left[ l_{(\alpha )}+n_{r}^{\prime
}\right] ^{-2};\text{ }n_{r}^{\prime }=1,2,3...\text{,}  \label{46}
\end{equation}
where $n_{r}^{\prime }=1+n_{r}$. Note that the energy levels become more and
more spaced as $\alpha $ tends to $1$, which means that the shift in the
energy levels due to the presence of the cosmic string increases with the
increasing of the angular deficit.

From the expression for the energy given by Eq. (\ref{46}) we can notice that
the levels without a $z$-component of the angular momentum are not shifted
relative to the Minkowski case. Except these levels all the other ones are
degenerated.

As an estimation of the effect of the cosmic string on the energy shift, let
us consider $\alpha \cong 0.999999,$ which corresponds to GUT cosmic
strings.
In this case the energy of the particle in the presence of the
cosmic string decreases of about $4\times 10^{-3}\%$ as compared with the
flat space-time value.

\vskip 1.0 cm

\centerline{\bf{III. Harmonic oscillator in the space-time of a cosmic string}}
%begin{figure}
%\special{center , \the\hsize 6cm 6cm 6cm , imagem.gif}
%\vspace{7cm}
%\end{figure}

The line element corresponding to the cosmic string space-time is given in
cylindrical coordinates by\cite{17}

\begin{equation}
ds^2=-dt^2+d\rho ^2+\alpha ^2\rho ^2d\theta ^2+dz^2,  \label{1}
\end{equation}
where $\rho \geq 0$ and $0\leq \theta \leq 2\pi $. The string is
 situated on the $%
z$-axis. In the special case $\alpha =1$ we obtain the
 Minkowski space in
cylindrical coordinates. This metric has a cone-like
 singularity at $\rho =0$%
. In other words, the curvature tensor of the metric (\ref{1}), considered
as a distribution, is of the form
\begin{equation}
R_{\text{ }12}^{12}=2\pi \frac{\alpha -1}\alpha \delta ^2(\rho ),  \label{a2}
\end{equation}
where $\delta ^2(\rho )$ is the two-dimensional Dirac $\delta $-function.

Now, let us consider the Schr\"{o}dinger equation in the metric (\ref{1})
which is given by
\begin{equation}
-\frac{\hbar ^{2}}{2\mu }\left[ \partial _{\rho }^{2}+\frac{1}{\rho }%
\partial _{\rho }+\frac{1}{\alpha ^{2}\rho ^{2}}\partial _{\theta
}^{2}+\partial _{z}^{2}\right] \psi (t,\rho ,\theta ,z)+V(\rho ,z)\psi
(t,\rho ,\theta ,z)=i\hbar \frac{\partial }{\partial t}\psi (t,\rho ,\theta
,z),  \label{a3}
\end{equation}
where $V(\rho ,z)$ is the interaction potential corresponding to a
three-dimensional harmonic oscillator which is assumed to be
\begin{equation}
V(\rho ,z)=\frac{1}{2}\mu w^{2}\left( \rho ^{2}+z^{2}\right) .  \label{2}
\end{equation}

We will now determine the eigenfunctions of the Schr\"{o}dinger equation
(Eq. (\ref{a3})) with the interaction potential given by
Eq. (\ref{2}), by searching for solutions of
the form
\begin{equation}
\psi (t,\rho ,\theta ,z)=\frac{1}{\sqrt{2\pi }}e^{-i \frac{Et}{\hbar}
+im\theta }R(\rho)Z(z).  \label{a4}
\end{equation}

Equation (\ref{a3}) leads to two ordinary differential equations for $R(\rho
)$ and $Z(z)$ which are given by

\begin{equation}
-\frac{\hbar ^{2}}{2\mu }\left[ \frac{1}{R(\rho )}\frac{d ^{2}R(\rho )%
}{d \rho ^{2}}+\frac{1}{R(\rho )\rho }\frac{d R(\rho )}{%
d \rho }-\frac{m^{2}}{\alpha ^{2}\rho ^{2}}\right] +\frac{1}{2}\mu
w^{2}\rho ^{2}=\Omega  \label{9}
\end{equation}
and
\begin{equation}
-\frac{\hbar ^{2}}{2\mu Z(z)}\frac{d ^{2}Z(z)}{d z^{2}}+\frac{1%
}{2}\mu w^{2}z^{2}=\varepsilon _{z},  \label{10}
\end{equation}
where $\Omega $ is a separation constant and such that
\begin{equation}
\Omega +\varepsilon _{z}=E.  \label{11}
\end{equation}

Equation (\ref{10}) is the Schr\"{o}dinger equation for a particle in the
presence of one-dimensional harmonic oscillator potential, and then we have
the well-known results
\begin{equation}
\varepsilon _{z}=\left( n_{z}+\frac{1}{2}\right) \hbar w;\text{ }%
n_{z}=0,1,2,...,  \label{12}
\end{equation}
with
\begin{equation}
Z(z)=2^{-\frac{n_{z}}{2}}\left( n_{z}!\right) ^{-\frac{1}{2}}\left( \frac{%
\mu w}{\hbar \pi }\right) ^{\frac{1}{4}}e^{-\frac{\mu w}{2\hbar }%
z^{2}}H_{n_{z}}\left( \sqrt{\frac{\mu w}{\hbar }}z\right) ,  \label{a5}
\end{equation}
where $H_{n_{z}}$ is the Hermite Polynomial.

Now, let us look for solutions of
Eq. (\ref{9}). Its solution can be written as
\begin{equation}
R(\rho )=\exp \left( -\frac \tau 2\rho ^2\right) \rho ^{\frac{\left|
m\right| }\alpha }F(\rho ),  \label{13}
\end{equation}
where $\tau =\frac{mw}\hbar $ and
\begin{equation}
F(\rho )=_1F_1\left( \frac 12+\frac{\left| m\right| }{2\alpha }-\frac{\mu
\Omega }{2\hbar ^2\tau },\frac A2;\tau \rho ^2\right)  \label{15}
\end{equation}
is the degenerate hypergeometric function, with $A=1+2\frac{\left| m\right| }%
\alpha $.

In order to have normalizable wave-function, the series in Eq. (\ref{15})
must be a polynomial of degree $n_{\rho }$, and therefore
\begin{equation}
\frac{1}{2}+\frac{\left| m\right| }{2\alpha }-\frac{\mu \Omega }{2\hbar
^{2}\tau }=-n_{\rho };\text{ }n_{\rho }=0,1,2,....  \label{17}
\end{equation}
With this condition, we obtain the following energy eigenvalues
\begin{equation}
\Omega =\hbar w\left( 1+\frac{\left| m\right| }{\alpha }+2n_{\rho }\right) .
\label{18}
\end{equation}
If we substitute Eqs. (\ref{18}) and (\ref{12}) into (\ref{11}) we get,
finally, the energy eigenvalues
\begin{equation}
E_{N}=\hbar w\left( N+\frac{\left| m\right| }{\alpha }+\frac{3}{2}\right) ,
\label{19}
\end{equation}
where $N=2n_{\rho }+n_{z}.$

Therefore, the complete eigenfunctions are then given by
\begin{eqnarray}
\psi (t,\rho ,\theta ,z) &=&C_{Nm}e^{-iE_N\frac t\hbar }e^{-\frac \tau 2\rho
^2}\rho ^{\frac{\left| m\right| }\alpha }F_1\left( \frac 12+\frac{\left|
m\right| }{2\alpha }-\frac{\mu \Omega }{2\hbar ^2\tau },\frac A2;\tau \rho
^2\right)  \nonumber \\
&&e^{im\theta }2^{-\frac{n_z}2}\left( n_z!\right) ^{-\frac 12}\left( \frac{%
\mu w}{\hbar \pi }\right) ^{\frac 14}e^{-\frac{\mu w}{2\hbar }%
z^2}H_{n_z}\left( \sqrt{\frac{\mu w}\hbar }z\right) ,  \label{b2}
\end{eqnarray}
where $C_{Nm}$ is a normalization constant. It is worth calling
 attention to the fact that the
presence of the cosmic string breaks the degeneracy of the energy levels.

Note that in the limit $\alpha \rightarrow 1$ we get the results
corresponding to the three-dimensional harmonic
oscillator in flat space-time as it should be.

In the case under consideration the shift in the energy spectrum
between the first two
levels in this background increases of about $10^{-5}\%$ as compared with
the flat Minkowski space-time case.

Now, if we consider the spherical harmonic oscillator potential
$ V(r)=\frac{1}{2}\mu \omega ^2 r^2$, we can do similar
 calculations using the metric of the cosmic
 string in spherical coordinates, in which case we obtain
 the following expression for the energy spectra
\begin{equation}
E_{n,l(\alpha )}=\hbar w\left( l_{(\alpha )}+2n+\frac{3}{2}\right) ;\text{ }%
n=0,1,2,...,  \label{900}
\end{equation}
which is a result similar to the previous one given by Eq. (\ref{19}).

\vskip 1.0 cm

\centerline{\bf {IV. Kratzer potential in the space-time
 of a global monopole%
}}

The solution corresponding to a global monopole in a $O(3)$ broken symmetry
model has been investigated by Barriola and Vilenkin\cite{8}.

Far away from the global monopole core we can neglect the mass term and as a
consequence the main effects are produced by the solid deficit angle. The
respective metric in the Einstein theory of gravity can be written
as\cite{8}
\begin{equation}
ds^2=-dt^2+dr^2+b^2r^2\left( d\theta ^2+\sin ^2\theta d\varphi ^2\right) ,
\label{c3}
\end{equation}
where $b^2=1-8\pi G\eta ^2$, $\eta $ being the energy scale of symmetry
breaking and $ r$, $ \theta$ and $ \varphi$ are usual spherical coordinates.

Now, let us consider a particle interacting with a Kratzer potential given by

\begin{equation}
V(r)=-2D\left( \frac Ar-\frac 12\frac{A^2}{r^2}\right) ,  \label{a}
\end{equation}
where $A$ and $D$ are positive constants, and placed in
the background space-time given by metric (\ref{c3}).

In order to determine the energy spectrum let us write the Schr\"{o}dinger
equation in the background space-time of a global monopole. Then, we get
\begin{equation}
-\frac{\hbar ^2}{2\mu b^2r^2}\left[ 2rb^2\frac \partial {\partial r}+b^2r^2%
\frac{\partial ^2}{\partial r^2}-{\bf L}^2-2D\left( \frac Ar-\frac 12\frac{%
A^2}{r^2}\right) \right] \psi ({\bf r})=E\psi ({\bf r}),  \label{b}
\end{equation}
where ${\bf L}$ is the usual orbital angular momentum operator. We begin by
using the standard procedure for solving Eq. (\ref{b}) and assume
that the eigenfunctions can be written as

\begin{equation}
\psi _{m,l}({\bf r})=R_l(r)Y_l^m(\theta ,\varphi ).  \label{c}
\end{equation}
Substitution of Eq. (\ref{c}) into Eq. (\ref{b}) leads to
\begin{equation}
-\frac{\hbar ^2}{2\mu }\frac{d^2g_l(r)}{dr^2}-2D\left( \frac Ar-\frac 12%
\frac{A^2}{r^2}\right) g_l(r)+\frac{\hbar ^2}{2\mu }\frac{l(l+1)}{b^2r^2}%
g_l(r)=Eg_l(r),  \label{d}
\end{equation}
where $g_l(r)=rR_l(r).$

The solution of Eq. (\ref{d}) can be written as
\begin{equation}
g_l(r)=r^{\lambda _l}e^{-\bar{\beta}r}F_l(r)  \label{e}
\end{equation}
where
\begin{equation}
\left. \lambda _l=\frac 12+\frac 12\sqrt{1+4\left( \frac{2\mu DA^2}{\hbar ^2}%
+\frac{l\left( l+1\right) }{b^2}\right) }.\right.  \label{f}
\end{equation}
Substituting Eq. (\ref{e}) into Eq. (\ref{d}) and making use of Eqs. (\ref{g}%
) and (\ref{f}) we obtain the equation for $F(z)$
\begin{equation}
z\frac{d^2F(z)}{dz^2}+(2\lambda _l-z)\frac{dF(z)}{dz}-\left( \lambda _l-%
\frac{2mAD}{\bar{\beta}\hbar ^2}\right) F(z)=0,  \label{h}
\end{equation}
where $z=2\bar{\beta}r$.

The solution of this equation is the confluent hypergeometric function
 $_1F_1\left( \lambda _l-\frac{\gamma ^2}{\bar{\beta}A}%
,2\lambda _l\text{ ; }2\bar{\beta}r\right) $, with $\gamma ^2=
\frac{2\mu DA^2%
}{\hbar ^2}.$

Therefore, the solution for the radial function $g_l(r)$ is given by
\begin{equation}
g_l(r)=r^{\lambda _l}e^{-\bar{\beta}r}{}_1F_1\left( \lambda _l-\frac{\gamma
^2}{\bar{\beta}A},2\lambda _l\text{ ; }2\bar{\beta}r\right) .  \label{i}
\end{equation}

In order to make $g_l(r)$ vanishes for $r\rightarrow \infty $, the confluent
hypergeometric function may increase not faster than some power of $r$, that
is, the function must be a polynomial. Hence
\begin{equation}
\lambda _l-\frac{\gamma ^2}{\bar{\beta}A}=-\bar{n}_r,\text{ }\bar{n}%
_r=0,1,2,...\text{ }.  \label{y}
\end{equation}
With this condition we find that the eigenvalues are given by
\begin{equation}
E_{l,\bar{n}_r}=-\frac{\hbar ^2}{2\mu A^2}\gamma ^4\left( \bar{n}_r+\frac 12+%
\sqrt{\frac 14+\frac{l\left( l+1\right) }{b^2}+\gamma ^2}\right) ^{-2}
\label{l}
\end{equation}

From this expression, we see
that when $b=1$ we recover the well-known result corresponding to a
particle submitted to the Kratzer potential\cite{15} as it should be.

It is worth noticing from expression for the energy given by Eq. (\ref{l}) that
even in the case in which the $z-$component of the angular momentum vanishes
the energy level is shifted relative to the Minkowski case.

As an estimation of the effect of the global monopole on the energy
spectrum, let us consider a stable global monopole configuration for which $%
\eta =0.19m_p$, where $m_p$ is the Planck mass. In this situation the shift in
the energy spectrum between the first two levels in this
space-time decreases of about $82\%$ as compared with the Minkowski
space-time. For symmetry breaking at grand unification scale, the typical
value of 8$\pi G\eta ^2$ is around $10^{-6}$ and in this case
the energy shift decreases of about $1\%$.
\vskip 1.0 cm
\centerline{\bf{V. Morse potential in the space-time of a global monopole}}

Now, let us consider the Morse potential\cite{16} which is used
 to describe the vibrations of a diatomic molecule and reads as

\begin{equation}
V(r)=D\left( e^{2\beta x}-2e^{-\beta x}\right) ;\,\text{ }x=\frac{r-r_{0}}{%
r_{0}};\text{ }\beta >0.  \label{47}
\end{equation}

In order to calculate and analyze the modifications in the energy
spectrum by the presence of a global monopole, let us
consider a particular situation and expand this potential around $r=r_{0}$
(or $x=0$) which leads to

\begin{equation}
V(r^{\prime })\cong -D+\frac{1}{2}Mw^{2}r^{\prime 2},  \label{50}
\end{equation}
where $r - r =r^{\prime}$ and the frequency $ \omega$ is defined
by $\frac{1}{2}Mw^{2}=\frac{%
D\beta ^{2}}{r_{0}^{2}}$.
This potential corresponds to the spherical harmonic oscillator
 potential minus a
constant.

We now write the Schr\"{o}dinger radial equation which is given by

\begin{equation}
\frac{1}{r^{\prime}}\frac{d}{dr^{\prime}}
\left( r^{\prime 2}\frac{dR}{dr^{\prime}}\right)-
\frac{M^{2}w^{2}r^{\prime 2}}{%
\hbar ^{2}}R(r^{\prime })-l\frac{\left( l+1\right) }{b^{2}r^{\prime 2}}%
R(r^{\prime })+\frac{2M}{\hbar ^{2}}E_{M}R(r^{\prime })=0.  \label{52}
\end{equation}
Therefore, if we put
\begin{equation}
R(r^{\prime })=\exp \left( -\frac{1}{2}M\frac{w}{\hbar }r^{\prime 2}\right)
r^{\prime^{ -\frac{1}{2}+\frac{1}{2}\sqrt{1+\frac{4}
{b^{2}}l(l+1)}}}F(r^{\prime
}),  \label{53}
\end{equation}
we find
\begin{equation}
r^{\prime }\frac{d^{2}F(r^{\prime })}{dr^{\prime 2}}+\left[ c-\frac{2Mw}{%
\hbar }r^{{\prime }2}\right] \frac{dF(r^{\prime })}{dr^{\prime }}+
\left[ P-\frac{2Mw}{%
\hbar }\right] r^{\prime }F(r^{\prime })=0,  \label{54}
\end{equation}
where the parameters $ c$ and $ P$ are given by
\[
c=1+\sqrt{1+\frac{4}{b^{2}}l\left( l+1\right) };\text{ }P=\frac{2ME_{M}}{%
\hbar ^{2}}-\frac{Mw}{\hbar }\sqrt{1+\frac{4}{b^{2}}l\left( l+1\right) }+%
\frac{2MD}{\hbar ^{2}}.
\]
By a change of variables,
\begin{equation}
Mwr^{\prime 2}=\rho ,  \label{tv1}
\end{equation}
Eq. (\ref{54}) is transformed into
\begin{equation}
\rho \frac{d^{2}F(\rho )}{d\rho ^{2}}+\frac{1}{2}\left[ c-2\frac{\rho }{%
\hbar }\right] \frac{dF(\rho )}{d\rho }+\left[ \frac{P}{4Mw}-\frac{1}{2\hbar
}\right] F(\rho )=0.  \label{55}
\end{equation}
Now, defining $\frac{\rho }{\hbar }=x,$ we get
\begin{equation}
x\frac{d^{2}F(x)}{dx^{2}}+\left[ \frac{c}{2}-x\right] \frac{dF(x)}{dx}%
+\left[ \frac{P\hbar }{4Mw}-\frac{1}{2}\right] F(x)=0,  \label{56}
\end{equation}
whose solution is
\begin{equation}
F(x)=_{1}F_{1}\left( \frac{1}{2}-\frac{\hbar P}{4Mw},\frac{c}{2};x\right) ,
\label{val}
\end{equation}
or, in terms of variable $r^{\prime }$, we have
\begin{equation}
F(r^{\prime })=_{1}F_{1}\left( \frac{1}{2}-\frac{E_{M}}{2\hbar w}-\frac{D}{%
2\hbar w}+\frac{1}{4}\sqrt{1+\frac{4}{b^{2}}l(l+1)},\frac{1}{2}+\frac{1}{2}%
\sqrt{\frac{4}{b^{2}}l(l+1)};\frac{Mwr^{\prime 2}}{\hbar }\right) .
\label{57}
\end{equation}
Note that this solution is divergent, unless
\begin{equation}
\frac{1}{2}-\frac{E_{M}}{2\hbar w}-\frac{D}{2\hbar w}+\frac{1}{4}\sqrt{1+%
\frac{4}{b^{2}}l(l+1)}=-n_{M};\,\,\text{ }n_{M}=0,1,2,3,...\, . \label{58}
\end{equation}
From Eq. (\ref{58}) we find the energy levels which are given by
\begin{equation}
E_{M}=\hbar w\left[ N_{lM}+\frac{3}{2}\right] -D,  \label{59}
\end{equation}
where
\begin{equation}
N_{lM}=\frac{1}{2}\left( \sqrt{1+\frac{4}{b^{2}}l\left( l+1\right) }%
-1\right) +2n_{M}.  \label{60}
\end{equation}

An estimation of the modifications in the energy spectrum in this
case show us a decrease in the energy eingenvalues of
$3\times 10^{-5}\%$ and $ 32\%$ in the cases of GUT and stable
monopoles, respectively, as compared to the flat Minkowski space-time case.

\vskip 2.0 cm

\section*{VI. FINAL REMARKS}

In the space-time of a cosmic string we studied the behavior of a particle
interacting with an harmonic oscillator and a Coulomb potential. The quantum
dynamics of such a single particle depends on the nontrivial topological
features of the cosmic string space-time. The presence of the defect shift
the energy levels in both cases as compared to the flat Minkowski
 space-time one. It is
interesting to observe that these shifts depend on the parameter that
defines the angle deficit and therefore arise from the global
features of the
space-time of a cosmic string.

In the case of the Kratzer and Morse  potentials in the space-time
of a global monopole the shifts in the energy levels are due to the combined
effects of the curvature and the nontrivial topology determined by the solid
deficit angle corresponding to the space-time of a global monopole.

The magnitude of the modifications in the spectra of the quantum
 mechanical  potentials by the presence of gravitational strings
 and monopoles are
all measurable, in principle. In fact, to produce observable
 modifications at the astrophysical scale, one needs an huge
number of particles in the states that we have studied. We can also
 have the possibility to get a very strong contribution
to real spectral effects if the topological defects (cosmic strings
and monopoles) occur in huge numbers, like in superfluid vortices in
neutron stars which are expected to occur with densities of $ 10^{6}$
vortices per square centimeter of stellar crossection.

Our estimation takes into account just a simple configuration with
 one cosmic string or one global monopole, but even in this simple
situation we believe that astrophysical observations concerning the
 detectability of cosmic strings and monopoles in the Universe can
 be made although it seems difficult to provide these verifications
 with the current observational detectability.

 An important application of these results
 could be found in the
astrophysical context in which these modifications enters in the
interpretation
of the spectroscopical data and these could be used as a probe of the
presence of a cosmic string or a global monopole in the Universe.

Finally, it is worth commenting that the study of a quantum system
in a nontrivial
gravitational background like the ones considered in this paper may
 shed some light on the problems of combining
quantum mechanics and general relativity.

\section*{Acknowledgments}

We acknowledge Conselho Nacional de Desenvolvimento Cient\'{\i }fico e
Tecnol\'{o}gico for partial financial support.

\end{document}